\documentclass[reprint,amsmath,amssymb,aps,pre]{revtex4-2}

\usepackage{hyperref}
\usepackage{url}
\usepackage{siunitx}
\usepackage{todonotes}
\begin{document}

\title{Assembly of Complex Colloidal Systems Using DNA}
\author{William M. Jacobs}
\affiliation{Department of Chemistry, Princeton University, Princeton, NJ, USA, 08544}
\email{wjacobs@princeton.edu}
\author{W. Benjamin Rogers}
\affiliation{Martin A. Fisher School of Physics, Brandeis University, Waltham, MA, USA, 02453}
\email{wrogers@brandeis.edu}

\begin{abstract}
Nearly thirty years after its inception, the field of DNA-programmed colloidal self-assembly has begun to realize its initial promise. In this review, we summarize recent developments in designing effective interactions and understanding the dynamic self-assembly pathways of DNA-coated nanoparticles and microparticles, as well as how these advances have propelled tremendous progress in crystal engineering. We also highlight exciting new directions showing that new classes of subunits combining nanoparticles with DNA origami can be used to engineer novel multicomponent assemblies, including structures with self-limiting, finite sizes. We conclude by providing an outlook on how recent theoretical advances focusing on the kinetics of self-assembly could usher in new materials-design opportunities, like the possibility of retrieving multiple distinct target structures from a single suspension or accessing new classes of materials that are stabilized by energy dissipation, mimicking self-assembly in living systems.  
\end{abstract}

\maketitle

\section{INTRODUCTION}
Self-assembly---the formation of ordered patterns or structures from initially disordered building blocks---is ubiquitous in the physical world.
Self-assembly takes place across a range of length scales in Nature~\cite{Whitesides2002Mar}, from the spontaneous freezing of water to form ice crystals, to the assembly of polypeptides to form amyloid fibrils or functional molecular machines, to the organization of bacteria into biofilms.
These remarkable examples have inspired scientists to experiment with designed structures that can similarly form via spontaneous self-assembly.
In fact, self-assembly stands out as one of the few practical strategies for making ensembles of nanometer-scale materials for nanotechnology and synthetic biology.
Tremendous strides in this direction have led to the development of robust techniques for assembling novel protein complexes with nanometer-scale precision~\cite{bale2016accurate}, carrying out computations with nucleic acids~\cite{seelig2006enzyme}, and fabricating complex metamaterials~\cite{tang2002spontaneous}, among many other achievements.
These efforts have also provided an ideal platform for testing theoretical frameworks derived from statistical physics, leading to a deeper understanding of entropy~\cite{manoharan2015colloidal,frenkel2015order} and rare events such as crystal nucleation~\cite{harland1997crystallization,auer2001prediction}.

For the purposes of this review, we define `self-assembly' to mean the spontaneous formation of ordered structures in mixtures of many discrete subunits as they relax toward equilibrium.
This definition is a special case of self-organization, which more generally implies the emergence of order from disorder, since we wish to focus on the creation of structures and materials with well defined spatial arrangements of their subunits.
In Sec.~\ref{sec:outlook}, we will broaden this definition to consider biologically inspired examples of non-equilibrium self-assembly, in which ordered structures are stabilized by continuous energy dissipation.

\subsection{Colloidal self-assembly}
Colloidal self-assembly describes scenarios in which the subunits have a characteristic size ranging from a few nanometers to a few micrometers (\SI{10}{nm}--\SI{1}{\mu m}).
As such, this class of subunits encompasses a diverse range of building blocks ranging from biological and synthetic macromolecules to inorganic nanoparticles and micrometer-scale polymer microspheres.
Colloidal self-assembly appears in many biological contexts, including canonical examples such as viral capsids and protein crystals, and is particularly relevant for the fabrication of nanotechnological devices.

A hallmark of self-assembly at colloidal length scales is that the interactions can be \textit{designed}.
Whereas the interactions between atoms or small molecules are fixed by physical laws, the shapes, chemical compositions, and microstructures of colloidal building blocks control their interactions in a user-specified manner.
For example, the interactions between proteins are governed by their amino-acid compositions and three-dimensional folds, while the interactions between polymer microspheres depend on their sizes and surface coatings.
Because interactions between colloidal subunits can be described via effective potentials that suitably average over an enormous number of molecular degrees of freedom~\cite{anderson2002insights}, a prevailing paradigm in colloidal self-assembly is to conceptualize the subunits as the fundamental objects, and to control their collective behavior by tuning their effective interactions~\cite{frenkel2002playing}.
Such `designability' is essential for realizing new materials via self-assembly.

\subsection{DNA-programmed colloidal assembly}
Approximately three decades ago, seminal papers by Chad Mirkin~\cite{mirkin1996dna} and Paul Alivisatos~\cite{alivisatos1996organization} proposed that effective colloidal interactions could be designed by leveraging the sequence specificity of DNA hybridization.
This paradigm, often called some version of `DNA-based programmable self-assembly', has since developed into a powerful and versatile strategy for directing colloidal self-assembly at the nanometer and micrometer scales.
The key idea is that by coating colloidal particles with short pieces of single-stranded DNA, one can design structures in which subunits bearing complementary DNA sequences selectively attract one another, providing the driving force for self-assembly.
Because DNA follows Watson--Crick base-pairing rules~\cite{watson1953molecular}, and because the free energy of DNA hybridization can be accurately predicted from the base sequences~\cite{SantaLucia2004}, DNA-guided interactions can be specified quantitatively.
In this way, one can `program' the assembly of complex structures, using a wide variety of particle materials, simply by selecting appropriate DNA sequences: the DNA acts like the `mortar' between the colloidal particle `bricks'.

In recent years, the field has taken large leaps toward realizing this vision.
In particular, the field has specialized in programming the structure of crystal lattices assembled from nanometer- and micrometer-scale particles~\cite{park2008dna,nykypanchuk2008dna,macfarlane2011nanoparticle,casey2012driving,rogers2015programming,wang2015crystallization}.
At the nanometer scale, the field has successfully assembled crystals with dozens of unique symmetries from one or two components~\cite{macfarlane2011nanoparticle,auyeung2014dna,liu2016diamond,lewis2020single,zhou2024space}, including quasicrystals~\cite{zhou2023colloidal}, crystals with large unit cells~\cite{lin2017clathrate}, and crystals that do not have atomic analogs~\cite{wang2022emergence}.
At the micrometer scale, the field has assembled a smaller diversity of crystal structures, but has nonetheless succeeded in exploiting the virtues of DNA-programmed assembly to build complex crystalline materials~\cite{rogers2015programming,wang2015crystallization,fang2020two,Hensley2022self}, including the diamond lattice~\cite{he2020colloidal} and other close relatives~\cite{wang2017colloidal,ducrot2017colloidal}.
Many of these advances have been described by excellent recent reviews in the field~\cite{jones2015programmable,rogers2016using,seeman2017dna,laramy2019crystal,kahn2022designer}.

\subsection{Scope of this review}
Despite this progress, many outstanding challenges and intellectual opportunities remain.
First, it remains difficult to assemble bulk crystalline materials with high yields or macroscopic dimensions, in large part because the kinetic pathways governing crystallization are less well understood than the equilibrium behavior of these crystals, and self-assembly is prone to kinetic trapping.
Second, efforts to target a wider variety of complex structures beyond binary crystals are just beginning.
Finally, the field has yet to realize the full potential of programmable self-assembly---that user-specified structures, dynamical behaviors, and even computations can be encoded in the interactions among subunits in highly multicomponent systems---at the colloidal scale.

This review summarizes recent progress and the current outlook for addressing these key challenges.
In particular, we focus on developments in understanding the dynamic pathways to assembly at the colloidal scale and the implications for optimizing experimental protocols.
We also highlight recent successes in directing the multicomponent assembly of supracrystalline and self-limiting architectures using DNA-programmed colloidal particles.
We conclude with a discussion of open theoretical questions, including how to design colloidal particles for efficient self-assembly kinetics, how to encode multiple target structures, and how to perform colloidal self-assembly under far-from-equilibrium conditions.
Progress in these directions promises to bring about the next wave of qualitative advances in DNA-programmed colloidal self-assembly.

\begin{figure*}
\includegraphics{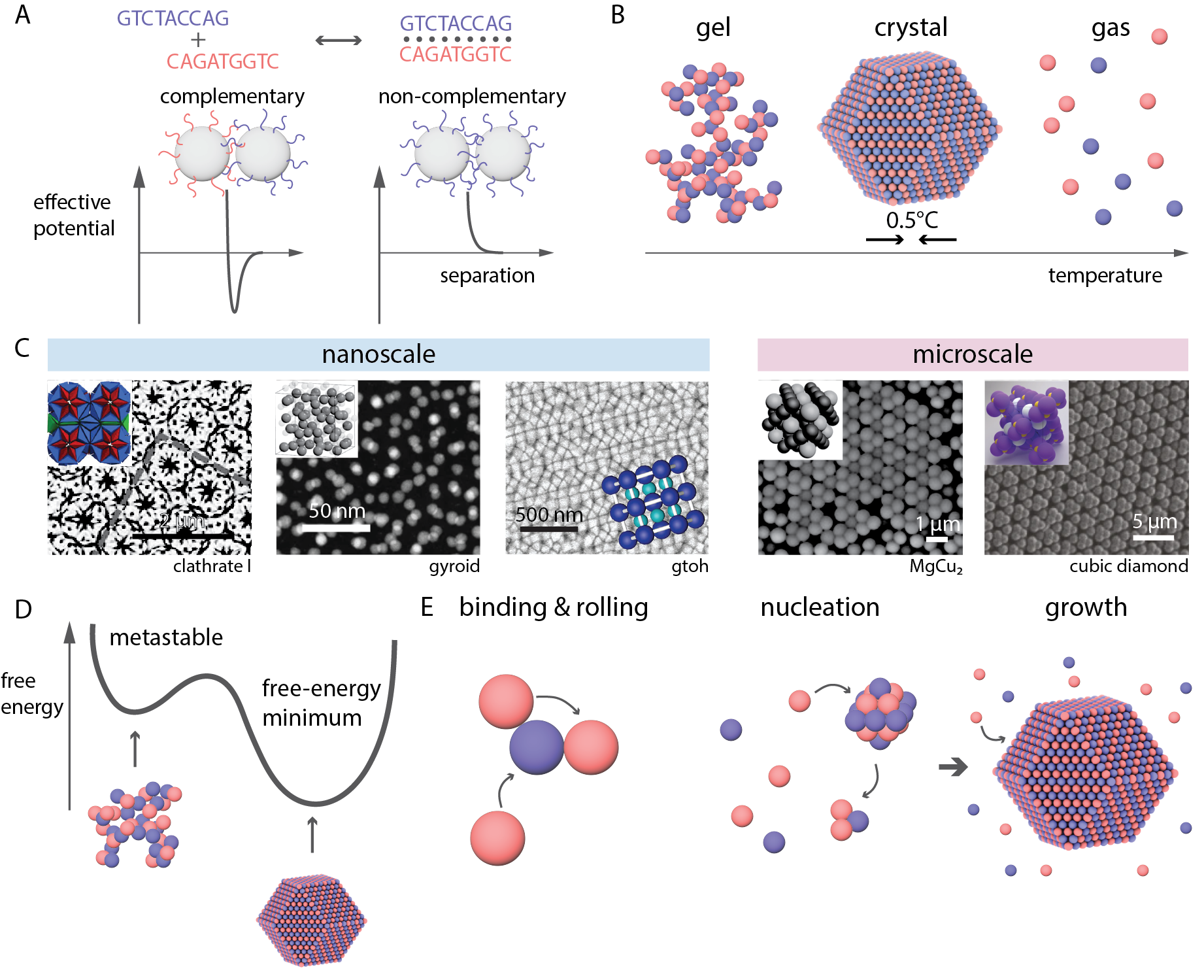}
\caption{An overview of DNA-programmable self-assembly. (A) The molecular recognition of Watson-Crick base pairing gives rise to sequence-specific interactions between DNA-grafted colloidal particles, characterized by effective interaction potentials. (B) In general, the cooperative interactions give rise to temperature-dependent behavior, limiting crystallization to a narrow temperature window. (C) Recent efforts to engineer more complex effective interaction potentials have yielded complex crystal phases. The outcome of self-assembly is sensitive to kinetic effects, like kinetic trapping (D), which depend sensitively on many dynamic factors, such as binding and rolling, nucleation, and growth (E). The clathrate I and gtoh crystal images in (C) were adapted from Refs.~\cite{lin2017clathrate,zhou2024space}, respectively. Reprinted with permission from AAAS. The gyroid, MgCu$_2$, and cubic diamond images in (C) were adapted with permission from Refs.~\cite{wang2022emergence,ducrot2017colloidal,he2020colloidal}, respectively.}
\label{fig1}
\end{figure*}

\section{EFFECTIVE INTERACTIONS AND EQUILIBRIUM PHASE BEHAVIOR}
The effective interactions between DNA-grafted particles arise from two dominant physical effects: 1) base pairing between complementary sequences induces an attraction that stabilizes contacts between particles bearing complementary base sequences; and 2) steric interactions and electrostatic repulsion between DNA molecules induce a repulsion that pushes particles apart (Fig.~\ref{fig1}A).  These two effects average over a large number of molecular degrees of freedom, resulting in a net effective interaction that is itself a temperature-dependent free energy. Because this topic has been covered extensively elsewhere~\cite{rogers2016using,angioletti2016theory}, we only emphasize the essential features and newest results that inform the rest of the review. 

DNA-mediated interactions can be largely grouped into two categories: direct binding and linker-mediated binding. In direct binding, DNA sequences coated on the particles' surfaces interact directly through base pairing. In linker-mediated binding, the DNA sequences grafted to the particles' surfaces are designed to be noncomplementary, so that the particles must be `linked' together via other DNA molecules dispersed in solution. Both strategies have been used since the inception of DNA-programmed colloidal assembly, and each presents unique practical and conceptual considerations.

\subsection{Measurements and models}
DNA-mediated effective interactions have been characterized in experiments and modeled quantitatively using a variety of approaches, including mean-field models~\cite{varilly2012general,angioletti2013communication,rogers2011direct,rogers2020mean} and coarse-grained computer simulations~\cite{li2013thermally,ding2014insights,mao2023regulating}. Of particular note, direct measurements of the effective pair-interaction potential using optical tweezers~\cite{biancaniello2005colloidal,rogers2011direct} and, more recently, total internal reflection microscopy~\cite{merminod2021avidity,Pine2022comprehensive} have revealed the full spatial and temperature dependence of the interactions. These measurements have played crucial roles in constraining the development of detailed theoretical models~\cite{biancaniello2005colloidal, rogers2012reply,mognetti2012predicting,Pine2022comprehensive}. The most recent such model~\cite{Pine2022comprehensive} builds on a decade-plus of development to provide a comprehensive view of how DNA hybridization, steric repulsion, van der Waals attraction, and other surface forces affect the effective interaction potential.

This combination of experiment and theory has revealed a wealth of understanding that is critically important for designing DNA-programmed colloidal self-assembly. First, the interactions are highly cooperative and therefore have an extremely strong temperature dependence.  In fact, the attractive interactions go from being comparable to $k_{\text{B}} T$ to being essentially irreversible over a few degrees Celsius, leading to very narrow temperature windows for crystallization (Fig.~\ref{fig1}B). Second, the range of attraction is comparable to the size of the grafted DNA molecules. Therefore, the range of attraction is comparable to the diameter of nanometer-scale particles, but is typically only $\sim1\%$ of the subunit diameter for micrometer-scale particles. The vast differences in the range of attractions between nanoparticles and microparticles lead to fundamental differences in their phase behavior and kinetics, including differences in the topology of the phase diagram and the dynamics of nucleation~\cite{anderson2002insights}. Third, the interactions are well-described by pair potentials, with some rare exceptions, as in the case where soluble linkers are mixed in small concentrations~\cite{lowensohn2019linker}. Finally, entropy plays a crucial role, giving rise to effective interactions that can have noncanonical dependencies on thermodynamic variables, like temperature and density, to yield unusual phase behavior like freezing upon heating~\cite{rogers2015programming} and other re-entrant transitions~\cite{lowensohn2019linker}.

Linker-mediated binding offers additional benefits over direct-binding interactions.
First, linker-mediated interactions offer an additional degree of freedom---the molar concentration of the DNA linkers---that can be used to adjust the effective interactions `on the fly'~\cite{xiong2009phase,lowensohn2019linker}. Second, linker-mediated binding has a finite working range of linker concentrations: At low concentrations, assembly is kinetically arrested~\cite{lowensohn2020self}, while at high concentrations, a disassembled fluid state is thermodynamically preferred over an assembled state even at low temperatures~\cite{lowensohn2019linker}. Lastly, linker-mediated binding is one of few practical strategies for potentially designing hundreds of interactions between dozens of unique components~\cite{lowensohn2019linker,rogers2020mean} owing to the fact that the interactions can be prescribed entirely via the design of the linker sequences. Simple estimates from experiment and theory suggest that this scheme could encode $>100$ interactions in a single mixture~\cite{lowensohn2019linker,rogers2020mean}, whereas typical experiments using direct binding utilize only a few specific interactions~\cite{macfarlane2011nanoparticle,wang2015crystallization,fang2020two}.

\subsection{Crystal engineering}
The deepening understanding of DNA-mediated effective interactions continues to propel the programmed assembly of equilibrium crystal structures forward, advancing from simple binary lattices in 2008~\cite{park2008dna,nykypanchuk2008dna} to wildly complex crystal lattices at the time of this review~\cite{lin2017clathrate,wang2022emergence,zhou2024space,ducrot2017colloidal,he2020colloidal} (Fig.~\ref{fig1}C).
These advances are driven in large part by the synthesis of subunits that go beyond completely isotropic pair-interaction potentials. At the nanometer scale, these developments have been enabled by using DNA-coated metal nanocrystals with polyhedral geometries~\cite{jones2010dna,lin2017clathrate,zhou2023colloidal,zhou2024space}, or by integrating other molecular engineering strategies like DNA origami~\cite{tian2016lattice,liu2016self,liu2016diamond,tian2020ordered}. At the micrometer scale, non-isotropic pair potentials have been enabled by the development of DNA-labeled `patchy particles'~\cite{ducrot2017colloidal,he2020colloidal}. At both scales, these new subunits give rise to preferential directional interactions that help to enforce certain crystal symmetries or stabilize particular crystal polymorphs, which are otherwise difficult to target using short-range isotropic interactions alone~\cite{mao2023regulating}. 

\subsection{Metamaterial applications}
The assembly of crystals with ever increasing complexity is just starting to be accompanied by the assembly of crystals with novel functionality. At the nanometer scale, it is now possible to assemble colloidal crystals with metamaterial properties, like shape memory~\cite{lee2022shape} and tunable mechanics~\cite{li2023ultrastrong}, and nanocrystal-based devices, like micromirrors~\cite{zornberg2023self} and three-dimensional arrays of Josephson junctions~\cite{shani2020dna}. 
At the micrometer length scale, recent studies demonstrate the assembly of crystals with optical metamaterial properties, like structural coloration in the visible~\cite{Hensley2022self,Hensley2023macroscopic} and a complete photonic bandgap in the near infrared~\cite{he2020colloidal}. These successes represent but a handful of the possibilities that can be achieved via the DNA-programmed self-assembly paradigm.  As we will articulate in the following sections, controlling the dynamical pathways and exploiting truly multicomponent assembly will open even more doors to metamaterial and device applications.

\section{DYNAMIC PATHWAYS}
While effective interactions are clearly an essential component of crystal engineering, they are not sufficient: Kinetic factors are equally important for programming colloidal self-assembly~\cite{whitelam2015statistical,jacobs2016self,gartner2022time}.
Indeed, it took twelve years from the conception of the idea~\cite{mirkin1996dna,alivisatos1996organization} to the assembly of the first crystals from nanoparticles~\cite{park2008dna,nykypanchuk2008dna}, while long-predicted crystalline phases are only recently being achieved in experiments~\cite{laramy2019crystal}.
A primary challenge has been the avoidance of kinetically arrested states, such as colloidal gels, that prevent equilibration to the target crystal structure (Fig.~\ref{fig1}D).
The crystallization of micrometer-scale particles has lagged even further behind, primarily owing to additional kinetic limitations that arise from the comparatively short range of their interactions.
These observations have motivated further study of the kinetic pathways governing crystallization in colloidal suspensions of DNA-grafted nanometer- and micrometer-scale particles (Fig.~\ref{fig1}E).

\subsection{Binding and rolling kinetics}
One essential kinetic consideration concerns particle binding and unbinding.
In this regard, effective interactions only tell half of the story, because, as free energies, they describe the ratios of these rates.
However, the absolute time scales turn out to be critical in practice.

At the micrometer scale, experimental measurements have revealed that high coating densities and many weak, transient molecular bridges enable rapid equilibration.
For example, binding/unbinding kinetics have been shown to be reaction-limited for DNA densities below roughly 2000 molecules per square micrometer~\cite{rogers2013kinetics}, while grafting densities above roughly 3000 molecules per square micrometer~\cite{wang2015crystallization} are required to crystallize micrometer-sized particles.
Similar stories emerge at the nanometer scale and for linker-mediated binding, where low linker concentrations lead to kinetic arrest and high linker concentrations enable crystallization~\cite{lowensohn2020self}.
As a result, the field has developed synthesis schemes to achieve high coating densities of DNA~\cite{hurst2006maximizing,zhang2013general,wang2015crystallization,wang2015synthetic,oh2019colloidal} and therefore fast kinetics~\cite{mao2020self}.

Rapid binding/unbinding kinetics are essential for self-assembly because they permit local rearrangements within growing assemblies.
Rearrangements are particularly important for crystallizing high-symmetry, isotropic building blocks like spheres, because the local crystal symmetry must emerge, collectively, to minimize the many-body free energy.
Consequently, an incoming particle needs to `hop' or `crawl' its way to its equilibrium position within the emerging lattice via trial and error~\cite{zheng2023hopping}.
A growing appreciation of these effects has led to the development of insightful models relating binding/unbinding kinetics to the rolling and sliding dynamics of colloidal motion~\cite{lee2018modeling,jana2019translational,marbach2022nanocaterpillar}.
Based on these insights, the field has moved toward using shorter DNA sticky ends and a greater reliance on precise temperature control to achieve reversible effective interactions~\cite{laramy2019crystal}.

\subsection{Nucleation}
\begin{figure*}
\includegraphics{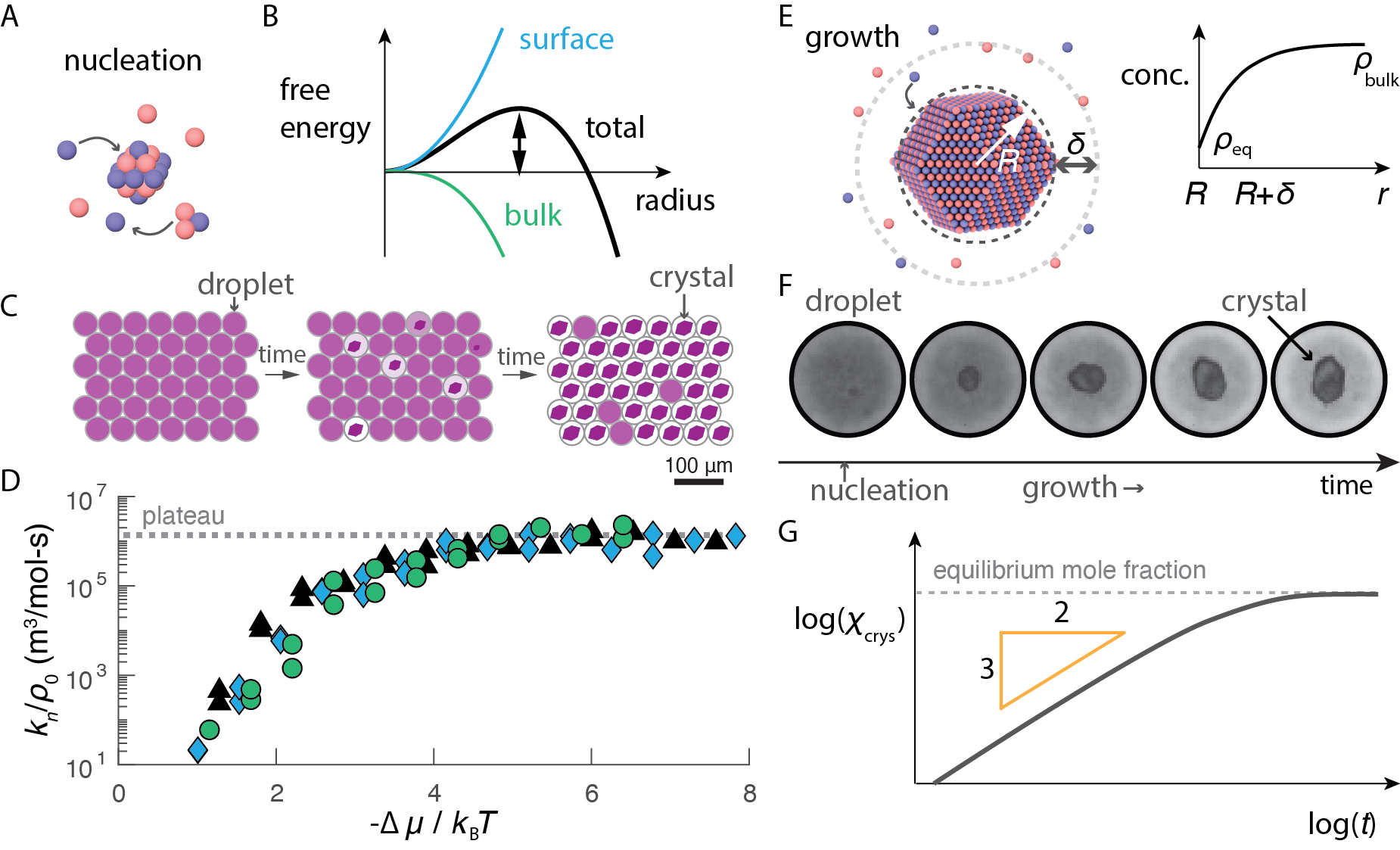}
\caption{Dynamic pathways to crystallization.
(A) Nucleation of a crystal from a fluid requires surmounting a free-energy barrier that results from a competition (B) between surface and bulk free energies. (C) The nucleation rate, $k_{\text{n}}$, can be measured by tracking the formation of crystals over time in many small droplets. (D) A plot of the rescaled nucleation rate versus the chemical potential difference, $\Delta\mu$, between the crystal and gas reveals that the nucleation barrier for micrometer-scale colloidal crystallization is described by classical nucleation theory. $\rho_0$ is the total colloid number density. (E) A growing crystal depletes monomers in its vicinity, leading to a spatially nonuniform concentration field, which varies radially from the equilibrium density, $\rho_\textrm{eq}$, to the bulk density, $\rho_\textrm{bulk}$, over a characteristic distance, $\delta$. (F) Optical micrographs showing the growth of a single colloidal crystal inside of a small droplet over time post nucleation. The droplet is roughly 60 micrometers in diameter. (G) A plot of the crystal mole fraction, $\chi_\textrm{crys}$, versus time, $t$, exhibits a power-law consistent with diffusion-limited growth. Panels (D) and (F) are adapted from Ref.~\cite{Hensley2022self}. Panel (C) is adapted with permission from Ref.~\cite{Hensley2023macroscopic}.}
\label{fig2}
\end{figure*}

Colloidal crystallization typically follows a dynamic pathway governed by nucleation and growth.
The nucleation step---the formation of the first transiently stable cluster that can continue to grow (Fig.~\ref{fig2}A)---necessitates surmounting a free-energy barrier that arises from a competition between interfacial and bulk free energies (Fig.~\ref{fig2}B).
Despite the conceptual simplicity of this pathway, studying nucleation quantitatively has proven to be a slippery problem.
Nucleation involves physical processes occurring over a wide range of time scales and is horribly sensitive to nonidealities common to experimental systems, including impurities, flow, and the presence of boundaries~\cite{gasser2001real,auer2001prediction}.
Moreover, crystals can alternatively emerge via homogeneous nucleation from a uniform solution, heterogeneous nucleation at an interface, or multistage nucleation, in which a cascade of transitions leads to the final crystalline phase~\cite{karthika2016review}.

\subsubsection{Homogeneous nucleation}
Recent quantitative studies of nucleation using micrometer-scale DNA-coated particles have begun to shed light on the dynamical complexities associated with DNA-programmed crystallization.
By monitoring the self-assembly of isolated colloidal crystals confined in hundreds of monodisperse, nanoliter-sized droplets (Fig.~\ref{fig2}C), direct measurements of the nucleation rate, $k_{\text{n}}$, have been made as a function of temperature, $T$, and the initial colloid density, $\rho_0$~\cite{Hensley2022self}.
Because the initial supersaturation can also be inferred at each condition, these data present a complete, quantitative picture of the nucleation kinetics and thus represent one of the most well controlled experimental tests of colloidal nucleation to date (Fig.~\ref{fig2}D)~\cite{Hensley2022self}.

These experiments reveal a number of important insights into the dynamics of crystallization.
First, they show that crystallization can proceed via homogeneous nucleation and that the dependence of the height of the free-energy barrier on the concentration, temperature, and interfacial tension can be described quantitatively by classical nucleation theory~\cite{oxtoby1992homogeneous}.
Second, by combining experiment and theory, this study shows that the absolute rate of nucleation and its concentration dependence are fundamentally altered by the effective friction between DNA-coated particles that arises from transient hybridization.
This effect is a direct consequence of the rolling and sliding kinetics mentioned above, which are an inherent feature of DNA-mediated interactions that distinguish colloidal interactions from simply scaled up atomic interactions.
Third, the nucleation rate varies by roughly one-million-fold over a temperature change of a quarter of a degree, thereby limiting the working temperature range for crystallization to about one tenth of one degree Celsius.
These findings help to rationalize why the field of DNA-programmed assembly struggled to form crystals of micrometer-scale particles for so long.

Going forward, the field would benefit greatly from similar experimental studies of the nucleation of different crystal symmetries and differently sized colloidal particles.
Perhaps such experiments could help to explain why some crystal structures fail to form at the micrometer scale, despite being thermodynamically favored~\cite{tkachenko2016generic}.
Theoretical predictions also suggest that nucleation pathways at the nanoscale and microscale could differ qualitatively owing to differences in the range of the effective interactions~\cite{anderson2002insights}, but these predictions have yet to be tested in experiment.
Importantly for the feasibility of nanoscale experiments, the methodologies developed in Ref.~\cite{Hensley2022self} are amenable to the study of nanoparticle crystallization since they do not rely on single-particle resolution.

\subsubsection{Non-classical nucleation}
Interestingly, not all colloidal crystals assembled from DNA-grafted colloidal particles form via a single nucleation step.
For example, recent experiments and simulations investigating dynamic crystallization pathways in two dimensions showed that nucleation can follow one-step pathways, as described above, or two-step pathways, in which one crystal phase forms first and then transforms into a different crystal symmetry~\cite{pretti2019size,fang2020two}.
What is particularly striking about these findings is that they were observed in arguably one of the simplest realizations of DNA-programmable self-assembly: a binary mixture of same-sized particles.
By exploring the attraction between like particles, these two independent studies found a rich diversity of crystal phases, including crystals with different symmetries, compositional orders, and stoichiometries~\cite{pretti2019size,fang2020two}.
Using computer simulations, both studies showed that the two-step crystallization pathway was driven entirely by competing thermodynamic forces between two crystal polymorphs with different interracial and bulk free energies~\cite{pretti2019size,fang2020two}.
This pathway is notable in that it is driven entirely by a diffusionless transformation within a single crystalline domain in coexistence with a dilute gas phase, as opposed to an intermediate dense fluid phase or a wetting layer at the crystal--gas interface.

An important takeaway is that by tuning the effective interactions, DNA-programmed colloids can access a diversity of dynamical assembly pathways in addition to the structural diversity of possible crystals.
This insight is likely to become increasingly relevant as the field pushes forward toward assembling more complex structures from increasingly complex mixtures of colloidal particles.
Moreover, the possibility for systems to undergo crystal-crystal transitions opens up the possibility of intentionally engineering structures that are kinetically inaccessible via single-step nucleation.

\subsection{Growth}
In addition to influencing nucleation, binding/unbinding kinetics also affect the growth rates of DNA-programmed crystals.
For example, slow rates of local rearrangements within a growing assembly can result in reaction-limited crystal growth, whereby the incorporation of colloidal particles into the nascent lattice is slower than the diffusive transport of colloidal particles to the crystal interface.
Recent experiments of micrometer-scale particles have shown that growth can be reaction-limited immediately following the formation of a critical nucleus, up to the point at which the crystal contains a few hundred particles~\cite{Hensley2022self}.

However, at later times, the growth of larger crystals is diffusion-limited.
The transition from reaction- to diffusion-limited growth occurs when the depletion of particles in the gas phase surrounding the growing crystal causes diffusion to become the rate-limiting process (Fig.~\ref{fig2}E).
In this regime, the details of binding/unbinding kinetics are less important, and growth becomes essentially deterministic, following a predicted power-law scaling with time~\cite{Hensley2022self,Hensley2023macroscopic} (Fig.~\ref{fig2}F--G).
An important consequence of deterministic dynamics is that crystal sizes can be precisely controlled, assuming that the concentration fields surrounding growing crystals do not substantially interfere with one another~\cite{Hensley2023macroscopic}.

\subsection{Protocol optimization}
\begin{figure*}
\includegraphics{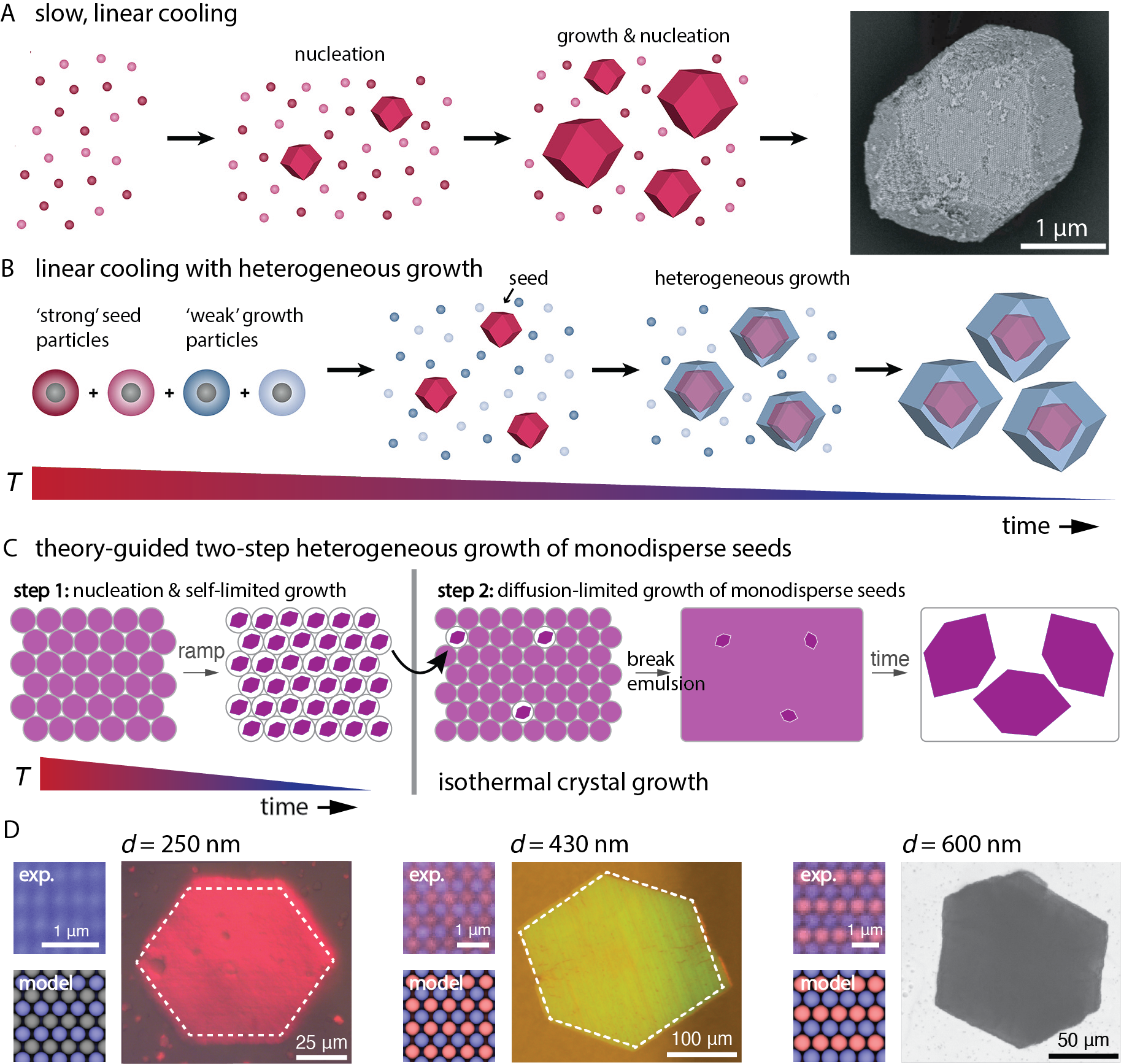}
\caption{Optimized assembly protocols yield macroscopic single crystals. (A) Slowly decreasing the temperature at a constant rate yields single crystals by favoring crystallization under conditions at which nucleation is rare yet growth can proceed. (B) Combining strong `seed' particles and weak `growth' particles promotes heterogeneous growth, limiting the size dispersity. (C) A theory-driven two-step protocol, in which monodisperse single crystals are assembled in droplets and then used to seed heterogeneous growth in a second isothermal step. (D) The two-step protocol in (C) can be applied to particles of differing diameters, $d$, to make macroscopic single crystals with different symmetries and lattice constants. (A) and (B) are reprinted (adapted) with permission from from Ref.~\cite{landy2023programming}. Copyright 2023 American Chemical Society. The micrograph in (A) is adapted with permission from Ref.~\cite{auyeung2014dna}. (C,D) are adapted with permission from Ref.~\cite{Hensley2023macroscopic}.}
\label{fig3}
\end{figure*}

The fundamental challenges imposed by the extreme temperature dependence of the nucleation and growth rates necessitate the design of new protocols for steering crystallization.
Inspired by the industrial processing of other crystalline materials, proposed approaches include the use of crystallization seeds to circumvent nucleation and nonequilibrium thermal annealing protocols.
Indeed, many of the initial successes in assembling colloidal crystals from DNA-grafted nanoparticles and microparticles used some form of nonisothermal protocol~\cite{park2008dna,nykypanchuk2008dna,casey2012driving}, but these empirical protocols were largely designed by intuition.
A similar approach used a slow linear cooling ramp to assemble the first single crystals of DNA-coated nanoparticles with well-defined crystal habits (Fig.~\ref{fig3}A), albeit with a broad range of crystal sizes~\cite{auyeung2014dna}.
Mirkin and co-workers later refined this technique to narrow the size dispersity by including `strong' seed particles and `weak' growth particles to separate the temperatures at which nucleation and growth take place along a cooling ramp~\cite{landy2023programming} (Fig.~\ref{fig3}B). 

The recent quantitative understanding of nucleation and growth kinetics now enable the rational design of near-optimal crystallization protocols.
In other words, it is now possible to optimize the rate of a thermal annealing protocol to yield single crystals of a target size in the shortest amount of time.
For example, quantitative theoretical models were used to predict optimal annealing rates for obtaining single crystals from micrometer-sized particles in microfluidic droplets~\cite{Hensley2022self}.
However, further experimental validation of this model~\cite{Hensley2023macroscopic} confirmed that fundamental physical constraints, owing ultimately to the strong temperature dependence of the binding/unbinding kinetics, limit the size of single crystals that can be assembled in confined environments.
This realization inspired the development of a two-step crystallization protocol~\cite{Hensley2023macroscopic} (Fig.~\ref{fig3}C), in which monodisperse crystallites assembled via an optimal annealing protocol in microfluidic droplets are then used to seed isothermal growth in a bath of additional particles.
Importantly, this protocol results in the assembly of macroscopic single crystals that are monodisperse in size due to the monodispersity of the microfluidic droplets in step one and the deterministic nature of diffusion-limited growth in step two.
This two-step protocol has been applied to assemble photonic single crystals with different symmetries and lattice constants, reaching sizes that are visible to the naked eye (Fig.~\ref{fig3}D).

\section{MULTICOMPONENT ASSEMBLY TO ACCESS NEW STRUCTURES}
The preceding two sections highlight the versatility and increasing robustness of DNA-programmed self-assembly as a tool for crystallizing colloidal particles. However, nearly all the colloidal crystals mentioned above are composed of only one or two components, owing to the fact that the effective interactions are short-ranged and have roughly isotropic specificity. Moreover, the longest length scales over which the subunits are ordered are comparable to the particle sizes. Going beyond these limitations to realize the full promise of programmable self-assembly requires the design of new subunits with additional user-specified attributes, including programmed directional interactions and subunit geometries.

In this section, we describe two particular assembly targets that differ conceptually from the colloidal crystals we have discussed so far: 1) supracrystalline materials that organize molecules or small particles at two or more distinct length scales; and 2) self-limiting architectures with one or more finite-size scales that can be arbitrarily large with respect to the subunits. These targets present compelling fundamental challenges in multicomponent design~\cite{jacobs2015self} and could lead to exotic material properties resulting, for example, from the coupling of plasmonic and photonic resonances that arise at different length scales~\cite{kravets2018plasmonic}. As in the preceding sections, we focus on recent efforts to assemble nanometer and micrometer-sized colloidal particles, including particles made by DNA origami~\cite{rothemund2006folding,wagenbauer2017we}, using DNA-programmed interactions.

\subsection{Supracrystalline assemblies}
\begin{figure*}
\includegraphics{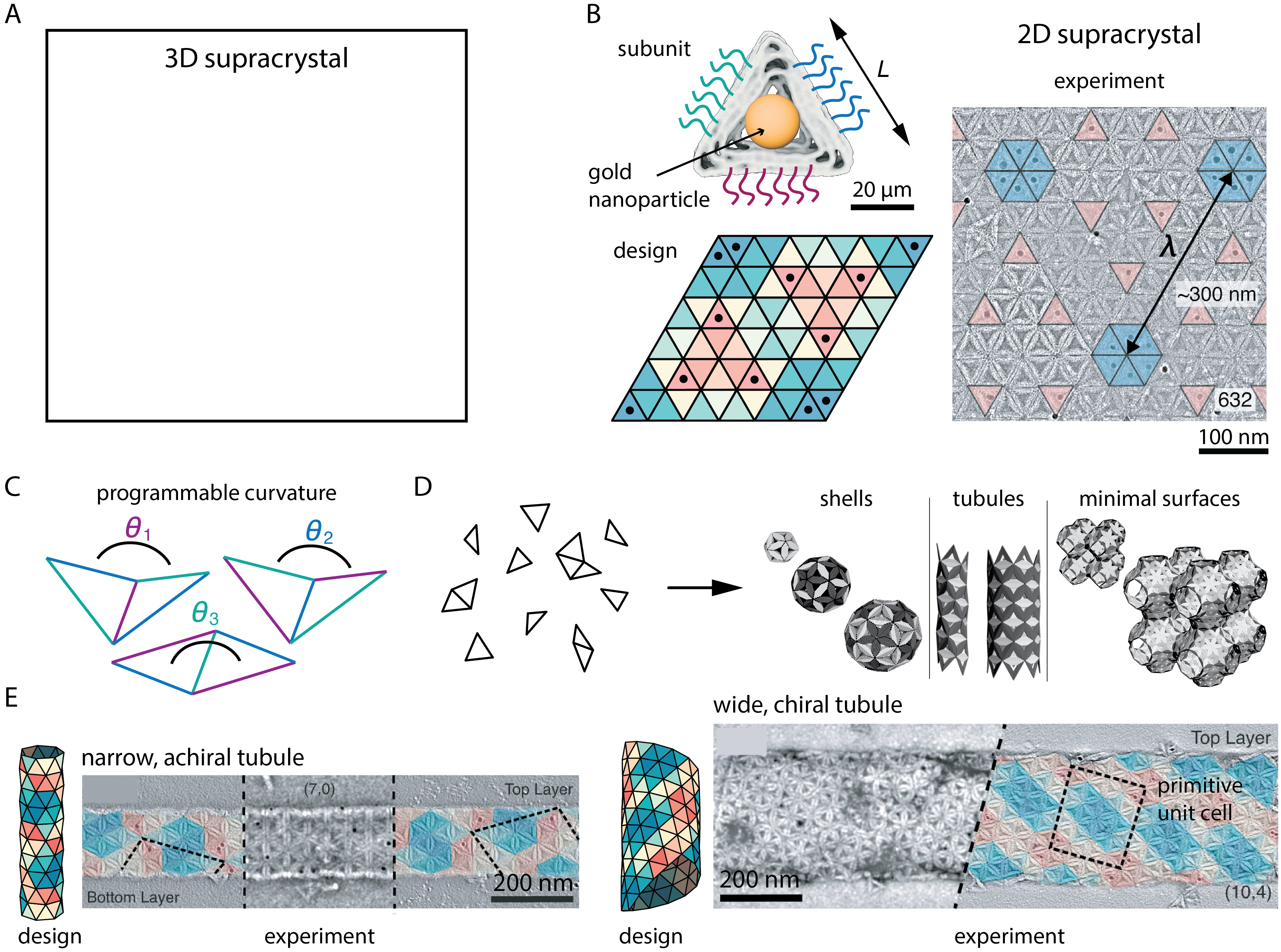}
\caption{Going beyond conventional crystalline assemblies to achieve longer length-scale features. (A,B) Examples of supracrystals in 3D (A) and 2D (B) with periodicities, $\lambda$, that are much larger than the subunit sizes, $\lambda \gg L$. Colors indicate unique components. The example in (B) belongs to the 632 Wallpaper group. The parallelogram shows the primitive unit cell. (C) By beveling the sides of the triangular subunit in (B), the subunits bind with user-specified dihedral angles, which give rise to self-closing structures, like the examples shown in (D). (E) Negative-stain EM images of cylindrical tubules: (left) an achiral, narrow tubule made from 14 components, shown in false color; (right) a chiral, wide tubule made from 16 components, shown in false color. EM micrographs of the `top' and `bottom' layers come from tomography. Dashed lines show the primitive unit cells. (B) and (E) are adapted with permission from Refs.~\cite{hayakawa2024,videbaek2023economical}, respectively.}
\label{fig4}
\end{figure*}

A new direction in colloidal assembly is to program crystalline materials with lattice parameters that are arbitrarily large with respect to the subunit size and with symmetries that are decoupled from the symmetries of the subunits themselves.
Such assemblies, which we refer to as `supracrystalline', are defined by multicomponent unit cells that can be used to organize different types of subunits.
Inspired by successful implementations of this concept using single-stranded DNA `bricks'~\cite{ke2014dna}, colloidal-scale subunits with valence-limited, directional interactions have been developed using DNA origami~\cite{rothemund2006folding,wagenbauer2017we}.
Specific interactions are encoded by the placement and sequence of DNA sticky ends that extend from the subunits (Fig.~\ref{fig4}A,B).
In this way, it is possible to make multicomponent libraries of subunits, of which only a small fraction are labeled with colloidal nanoparticles using the specificity of DNA hybridization.
These libraries can then be used to create crystals with a periodicity, $\lambda$, that is specified by the multicomponent unit cell as opposed to the size of the nanoparticle, $L$ (Fig.~\ref{fig4}A,B)

One of the most interesting current challenges in this direction is the development of economical design principles for programming the interactions within a multicomponent crystal. Whereas in the case of DNA bricks each multicomponent unit cell can feasibly be assembled from one copy of each component (i.e., the `fully-addressable' limit)~\cite{ke2014dna}, this approach is intractable using DNA-origami subunits, which must be folded and purified separately. In the colloidal case, it might therefore be beneficial to maximize the `subunit economy', defined as the size of the unit cell divided by the number of unique components required to create it~\cite{hayakawa2024}. While it is simple to pose this design challenge in principle---i.e., what is the minimum number of components required to build a target unit cell of a given size and what effective interactions are required to specify its assembly?---solving this problem is challenging given the enormity of the design space of multicomponent unit cells.

Confronting this challenge requires new design strategies to identify economical subunit designs that assemble into user-specified multicomponent structures. One promising approach for tackling this problem, called `SAT assembly', solves for the interaction matrix specifying the assembly of a user-defined unit cell by transforming the combinatorial problem into a Boolean satisfiability problem (SAT)~\cite{romano2020designing,russo2022sat,pinto2023design}. This approach has been applied to the design of complex multicomponent crystals, including crystals in the diamond family~\cite{liu2023inverse}. A second approach exploits the symmetries inherent to multicomponent crystals. In this way, deterministic algorithms can be designed to generate interaction matrices specifying the assembly of a multicomponent two-dimensional tiling~\cite{hayakawa2024} or a three-dimensional crystal~\cite{kahn2022encoding}. One example of such a tiling is shown in Fig.~\ref{fig4}B, which exploits 2-, 3-, and 6-fold rotational symmetries to create a multicomponent unit cell containing 72 subunits using only 12 components. An advantage of this second approach is that it enables the design of multicomponent unit cells containing hundreds of components, which pose an intractable problem for approaches that require simulations to identify and eliminate off-target structures~\cite{bohlin2023designing}.

As in the case of conventional colloidal crystallization, future opportunities in this area lie in understanding the dynamics of supracrystal nucleation and growth. Here, existing theories provide several interesting predictions to be explored in experiment. For example, theoretical predictions show that the nucleation barriers and growth pathways for crystals consisting of many components can be very different from those of simple crystals~\cite{jacobs2015rational}. In particular, and in contrast with simple crystals, theory suggests that it is possible to tune the kinetics of nucleation and growth separately, thus minimizing defect formation. Therefore, optimizing multicomponent crystals in more complex scenarios may require trading economy of subunit types for improved self-assembly kinetics.

\subsection{Self-limiting assembly via self-closure}
Designing self-limiting structures that have a least one finite dimension that is arbitrarily large with respect to the subunits, yet are still programmed through information encoded in the subunits themselves, presents further interesting opportunities and challenges beyond programming crystallization alone. 
Unlike spatially unlimited crystals, self-limiting structures need some way to `measure' their own size during growth in order to terminate assembly at a finite size.
Fully-addressable assembly, as demonstrated using single-stranded DNA bricks~\cite{ke2012three,ke2014dna,ong2017programmable}, provides one approach to self-limitation, in which case a finite-size structure is built from exactly one copy of each component. However, because this approach requires an enormous number of distinct subunit types, fully-addressable assembly is impractical for assembling DNA-origami building blocks when the target finite-size structures are much larger than just a few subunits.

Self-closing assembly, in which the local curvature defines the finite global size, has recently been explored as a more economical alternative to creating self-limiting architectures from DNA-origami building blocks. The basic idea is to program the building-block geometry to direct the growing assembly to close upon itself. By changing the local curvature and the interaction specificity through the design of the building blocks (Fig.~\ref{fig4}C), it is possible to target structures with varying finite sizes, symmetries, and topologies, including sizes that can be many times larger than the size of the subunits (Fig.~\ref{fig4}D)~\cite{Hagan2021equilibrium,duque2023limits}. Recent striking examples in colloidal self-assembly include the formation of rings~\cite{Wagenbauer2017gigadalton}, icosahedral capsids~\cite{Sigl2021programmable,wei2024hierarchical} and cylindrical tubules with programmable diameters, helicity, and handedness (Fig.~\ref{fig4}E)~\cite{Hayakawa2022,videbaek2023economical,karfusehr2024self}. 

Yet again, kinetic factors have been found to play a dominant role in determining the structures that form via self-closing assembly. In particular, growing assemblies typically become kinetically arrested at the moment of closure, owing to the formation of a large number of subunit-subunit contacts and an insurmountable free-energy barrier to reopening. Although ordered structures with few defects are able to form in this manner (Fig.~\ref{fig4}E), the distribution of states (i.e., tubules with slightly different diameters and helicity) is unable to relax to the equilibrium distribution~\cite{Fang2022,Videbaek2022}. The polymorphism that results from the kinetic arrest of thermal bending fluctuations worsens with increasing target size, making it increasingly difficult to accurately program the self-limiting dimension.

As in the case of supracrystalline lattices, trading subunit economy for the increased specificity of multicomponent assembly can address this design challenge.
In the case of cylindrical tubules, a target structure can be conceptualized as a triangulated sheet that closes upon itself by connecting two points on the growing sheet, which we refer to as `vertices' (Fig.~\ref{fig4}D).
While a specific tubule state may be preferred by the assembly's programmed curvature, the finite bending rigidity of the sheet permits the formation of many structurally similar tubules with similar bending energies.
These accessible geometries can be thought of as off-target states that occupy an area of vertices around the target-state vertex.
By assembling the sheet from an increasing number of distinct components, many of these off-target states become disallowed, as they do not satisfy the subunit binding rules imposed by the programmed subunit interactions.
Consequently, the number of off-target structures that are accessible via thermal fluctuations decreases.
As first predicted by theory and simulation~\cite{Videbaek2022,Fang2022}, and recently validated in experiment~\cite{videbaek2023economical}, only the target geometry forms when the size of the multicomponent unit cell exceeds the area of vertices that can be accessed by thermal fluctuations.
This criterion determines the minimum number of components required for achieving complete selectivity~\cite{videbaek2023economical}, highlighting the complementary roles that geometry and interaction specificity can play in programmable colloidal assembly.
Similar arguments apply for other self-closing architectures, like those illustrated in Fig.~\ref{fig4}D.

\section{OUTLOOK}
\label{sec:outlook}
Given these advances in experimental techniques and physical understanding, the field of DNA-programmed colloidal self-assembly is poised to tackle materials-design problems of even greater complexity.
However, achieving the full promise of programmable self-assembly requires a paradigm shift from viewing self-assembly kinetics as a complicating practical challenge to an exploitable design \textit{feature}.
This change of perspective necessitates fundamental theoretical advances and the development of conceptually new design strategies.
In this section, we highlight three emerging avenues toward this goal.

\subsection{Solving the `inverse problem' for dynamical pathways}
\begin{figure*}
\includegraphics{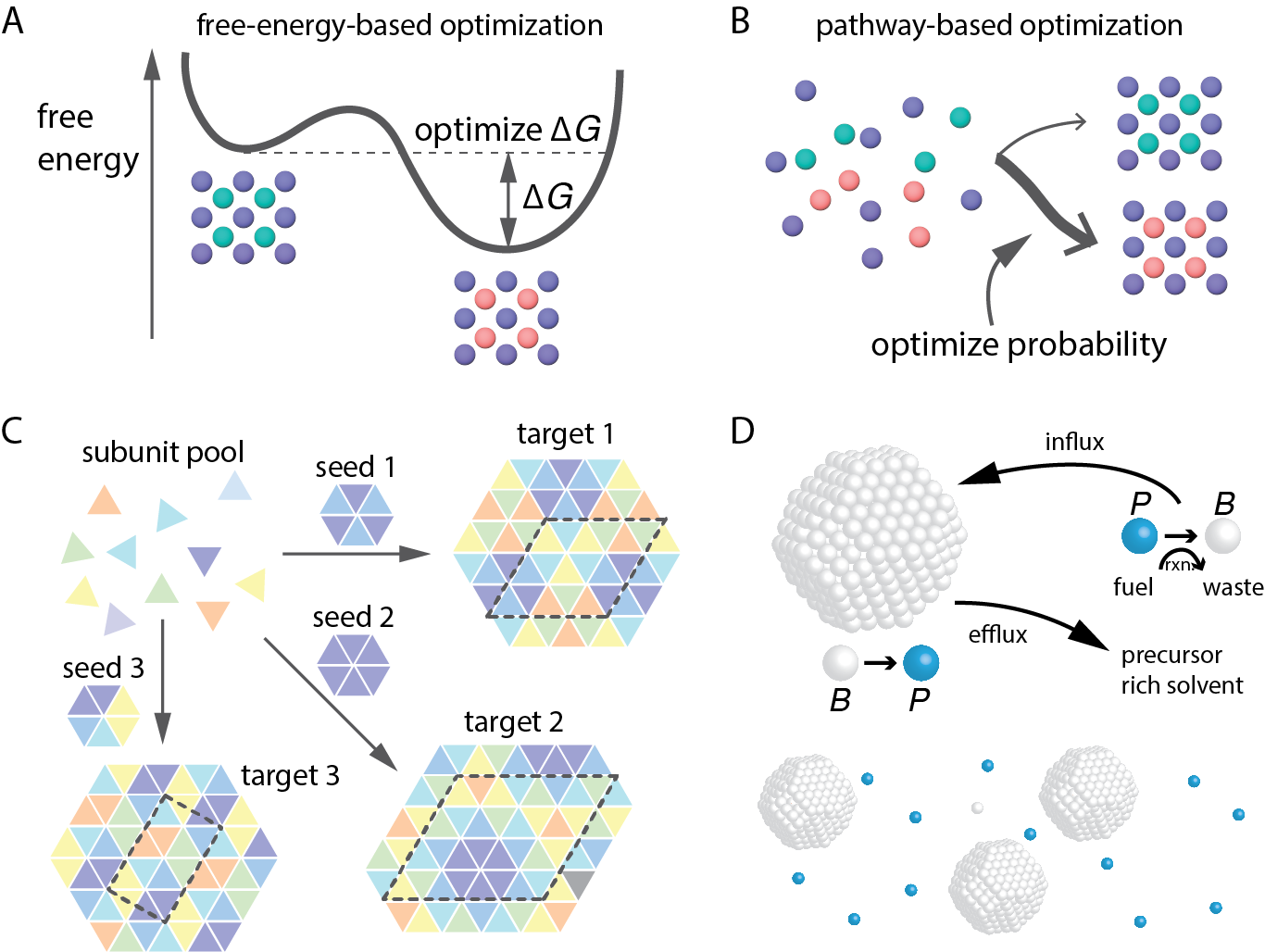}
\caption{Opportunities for designing colloidal self-assembly dynamics and kinetics. (A) Conventional inverse methods tune effective interactions to maximize the free-energy difference, $\Delta G$, between competing structures, whereas (B) pathway-based optimization aims to maximize the probability of assembling a target structure via a dynamical trajectory.  (C) Subunits in a multicomponent mixture can be designed to assemble into multiple distinct target structures, which can then be `retrieved' via specific nucleation seeds.  (D) Self-assembly at a nonequilbrium steady state exploits continuous energy dissipation, for example in the form of driven chemical reactions, to reach configurations that are unstable at equilibrium, such as ensembles of monodisperse finite-size structures.}
\label{fig5}
\end{figure*}

Colloidal self-assembly has long been viewed as an ideal playground for testing `inverse' statistical mechanics approaches, where the general goal is to design subunit interactions to achieve a user-specified target behavior~\cite{torquato2009inverse,miskin2016turning,sherman2020inverse}.
In the context of DNA-programmed self-assembly, this goal has traditionally meant designing an effective interaction potential, which could then be implemented by tuning colloidal particle properties, DNA sequences, and solution conditions.
Since the early days of this field, inverse approaches have primarily been used to find effective potentials for targeting crystal structures by optimizing free energy differences between competing crystal lattices~\cite{rechtsman2005optimized} (Fig.~\ref{fig5}A) or finite-size structures~\cite{Zeravcic2014size,halverson2013dna}.
Theory jumped out ahead of experiments in this regard, predicting the requirements for stabilizing the equilibrium crystal structures~\cite{tkachenko2002morphological,lukatsky2004phase,martinez2011design,scarlett2011mechanistic,tkachenko2016generic} that have only recently been fabricated in high yields.

Yet since kinetic considerations are now understood to be central to determining the outcome of an experiment, the inverse problem should be recast to target dynamical pathways directly in the design process.
By placing self-assembly kinetics on equal or greater footing relative to equilibrium free energy optimization, and thus considering the actual bottlenecks that arise in assembly experiments, inverse approaches are likely to provide better predictions for experiments, reducing the need for \textit{post hoc} optimization.
Furthermore, this approach opens up the possibility of targeting structures that are kinetically accessible but are not global free-energy minima.
These challenges represent the frontier of inverse statistical mechanics.

An essential first step is to understand the relationship between kinetic and thermodynamic optimality of effective colloidal interactions, since the accumulated experience of the field suggests that these objectives are often in conflict~\cite{villar2009self,hagan2006dynamic,jacobs2015rational,zenk2018optimizing}.
Current efforts to understand this trade-off systematically are utilizing modern multiobjective optimization methods.
For example, calculations of the Pareto front---i.e., the set of designs beyond which all objectives cannot be simultaneously improved---have shown that the trade-offs between kinetic and thermodynamic objectives depend on the target structure~\cite{trubiano2021thermodynamic} and the assembly conditions~\cite{chatterjee2024multi}, such as the initial particle concentration.
Advances in active-learning techniques for efficiently exploring high-dimensional parameter spaces~\cite{chatterjee2024multi} are enabling these calculations for complex and experimentally realistic simulation models.
This line of inquiry will ultimately allow us to understand when and why kinetic and thermodynamic objectives are in conflict, informing the development of improved design algorithms.

Inverse methods that target optimal self-assembly kinetics are already in development.
One strategy leverages classical theories of barrier-crossing events~\cite{miller2010exploiting,goodrich2021designing} or coarse-grained kinetic models~\cite{trubiano2022optimization} to optimize design parameters to maximize a finite-time yield or to minimize a first-passage time.
These methods generally require problem-specific intuition for defining relevant metastable states that an assembly pathway might traverse.
More recently, alternative approaches have been proposed to optimize molecular simulations of self-assembly directly (Fig.~\ref{fig5}B).
Promising methods utilize 1) automatic differentiation~\cite{goodrich2021designing,curatolo2022assembly}, a technique for computing gradients of simulation observables, 2) path reweighting~\cite{bolhuis2023optimizing}, in which an ensemble of simultaneous trajectories is interpreted probabilistically to compute the change in a first-passage time due to a change in the simulation parameters, or 3) reinforcement-learning~\cite{whitelam2021neuroevolutionary} and nonequilibrium-control~\cite{chennakesavalu2023unified} strategies, which can be used to optimize the effective interactions and time-dependent protocols based on kinetic objectives.
In all these cases, modern machine-learning tools are proving to be indispensable~\cite{sanchez2018inverse,dijkstra2021from} due to the inherent complexity of optimizing simulation trajectories or stochastic transitions as opposed to static configurations.
These approaches, while still in early stages of development, could be applied to design kinetic pathways for colloidal self-assembly by optimizing over a space of effective potentials that is accessible to experiments.

\subsection{Designing for multiple target structures}
Focusing design approaches on dynamical pathways opens up the possibility of assembling multiple structures from the same collection of colloidal building blocks.
In this case, the `retrieval' of a particular structure depends on the protocol, such as the solution conditions or the seed from which a structure is grown.
While alternative structures are always possible in complex systems---and indeed are the root cause of many kinetic pitfalls discussed in previous sections---the key distinction here is that a discrete set of desired structures is specified in the design process.
Ensuring the existence of a dynamical pathway to each target and the ability to control which pathway is taken then become the key objectives of the design process.
This paradigm could provide a route to multifunctional and stimulus-responsive colloidal materials, potentially incorporating logical operations that are possible with DNA computing~\cite{woods2019diverse}.

Within the realm of colloidal assembly, multiple distinct crystal structures have been assembled from a common pool of colloidal particles in a single experiment~\cite{rogers2015programming,lowensohn2020self}.
In these examples, the selection of one out of two or three possible crystals, differing in either their subunit compositions~\cite{rogers2015programming} or crystal symmetry groups~\cite{lowensohn2020self}, is determined either by varying the temperature~\cite{rogers2015programming} or concentrations of specific linker sequences~\cite{lowensohn2020self}.
Theoretical tools showing how to stabilize different structures at different temperatures in a multicomponent system~\cite{bupathy2022temperature} could be adapted to move forward along these lines.

Alternatively, different colloidal structures could in principle be selected by an initial seed at constant temperature and linker concentrations (Fig.~\ref{fig5}C).
In comparison to temperature control, this scheme allows for the encoding and subsequent controlled self-assembly of a much richer variety of structures, since the number of structures that can be suitably encoded is controlled by the number of subunit types~\cite{murugan2015multifarious,jacobs2021self} and the degree of interaction specificity~\cite{huntley2016information,chen2023emergence}.
The viability of this `multifarious' approach to self-assembly has been demonstrated experimentally using DNA tiles, where specific encoded structures were individually retrieved with high fidelity by providing appropriate nucleation seeds~\cite{evans2024pattern}.
DNA-programmed colloids provide an ideal platform for translating these ideas to the self-assembly of both finite-size and extended structures at the nanometer or micrometer scale.

\subsection{Designing nonequilibrium steady-states}
Another exciting direction in programmable assembly goes beyond the limitations of thermal equilibrium altogether.
Instead, `dissipative' assembly seeks to access structures that are not stable or kinetically accessible at thermal equilibrium~\cite{van2017dissipative,weber2019physics}.
Dissipative means that energy is continually supplied to the system, whether in mechanical, chemical, or thermal form, in order to maintain certain degrees of freedom away from equilibrium.
Examples of nonequilibrium assembly, such as microtubule growth~\cite{mitchison1984dynamic}, are commonly found in Nature, and to a much more limited extent in synthetic settings~\cite{spaeth2021molecular,nakashima2021active}.
Successfully translating these concepts to DNA-guided programmable assembly would mean that dynamical assembly pathways are no longer constrained to equilibrium mechanisms, allowing, for example, the implementation of proofreading~\cite{hopfield1974kinetic} during self-assembly~\cite{murugan2012speed,zhu2023proofreading}.
Dissipative colloidal assembly could also provide new routes to structures with unique structural properties, including self-limiting structures~\cite{zwicker2017growth,wurtz2018chemical}.
Finally, this approach could be used to imbue colloidal assemblies with dynamical properties that break time-reversal symmetry and are thus not possible at equilibrium.

A number of significant challenges must be surmounted to achieve these aims.
First, it is important to identify dissipative schemes that lead to behaviors and control parameters that differ qualitatively from what is possible at equilibrium.
To this end, emerging design rules for controlling collective behavior via driven chemical reactions~\cite{kirschbaum2021controlling,cho2023nucleation} could be adapted to DNA-programmed colloidal assembly to potentially create a suspension of many same-size assemblies (Fig.~\ref{fig5}D).
Second, new theories and design algorithms for optimizing colloidal interactions are required, since many standard statistical mechanical techniques for computing ensemble averages and gradients do not apply at a nonequilibrium steady state.
For example, numerically tractable schemes have been proposed to compute the minimum dissipation required to stabilize a nonequilibrium structure~\cite{nguyen2016design} and to optimize colloidal interactions at a nonequilibrium steady state~\cite{das2023nonequilibrium}.
Last but not least, practical experimental schemes must be developed to implement these ideas.
Tantalizing examples, such as dissipative assembly of linear structures using carefully tuned chemical reactions~\cite{spaeth2021molecular,nakashima2021active}, suggest that this goal may also be achievable using complex, fully programmable colloidal systems at nanometer and micrometer length scales.

\section{CONCLUSIONS}
In summary, recent breakthroughs in DNA-programmed colloidal self-assembly can largely be attributed to advances in designing effective interactions and developing a quantitative understanding of the dynamic pathways to crystallization.
Detailed measurements of the dynamics of nucleation and growth are guiding the development of optimal protocols for crystallization.
These advances are helping to rationalize and overcome kinetic bottlenecks, enabling the assembly of macroscopic single crystalline materials from DNA-programmed colloids.
In parallel, efforts to incorporate directional interactions and programmed subunit curvature are increasing the diversity of accessible crystal lattices, multicomponent supracrystalline materials, and self-limiting architectures.
These new directions are essential for realizing the full potential of DNA-programmed colloidal self-assembly.

Going forward, we anticipate that exciting discoveries will result from new approaches to programming the dynamics of self-assembly.
To this end, improved theoretical tools are needed to design effective interactions and assembly protocols on the basis of dynamical pathways as opposed to equilibrium free energies.
Kinetically controlled pathways could then be rationally designed to select specific targets from many encoded structures in multifarious mixtures, or to achieve even greater flexibility by performing colloidal self-assembly at nonequilibrium steady-states.
For these reasons, despite the fantastic achievements of this field over the past two decades, we strongly believe that the most important and exciting discoveries in the area of programmable self-assembly lie ahead.

\section*{DISCLOSURE STATEMENT}
The authors are not aware of any affiliations, memberships, funding, or financial holdings that
might be perceived as affecting the objectivity of this review. 

\section*{ACKNOWLEDGMENTS}
We acknowledge Thomas E. Videb\ae k and Rees F. Garmann for providing useful comments on an early draft of the review. WBR acknowledges funding from the National Science Foundation (DMR-2214590) and the Brandeis NSF MRSEC, Bioinspired Soft Materials, DMR-2011846. WMJ acknowledges funding from the Princeton NSF MRSEC, Princeton Center for Complex Materials, DMR-2011750.


\begin{thebibliography}{150}
\expandafter\ifx\csname natexlab\endcsname\relax\def\natexlab#1{#1}\fi

\bibitem{Whitesides2002Mar}
Whitesides GM, Grzybowski B. 2002.
\textit{Science} 295(5564):2418--2421

\bibitem{bale2016accurate}
Bale JB, Gonen S, Liu Y, Sheffler W, Ellis D, et~al. 2016.
\textit{Science} 353(6297):389--394

\bibitem{seelig2006enzyme}
Seelig G, Soloveichik D, Zhang DY, Winfree E. 2006.
\textit{Science} 314(5805):1585--1588

\bibitem{tang2002spontaneous}
Tang Z, Kotov NA, Giersig M. 2002.
\textit{Science} 297(5579):237--240

\bibitem{manoharan2015colloidal}
Manoharan VN. 2015.
\textit{Science} 349(6251):1253751

\bibitem{frenkel2015order}
Frenkel D. 2015.
\textit{Nature Materials} 14(1):9--12

\bibitem{harland1997crystallization}
Harland J, Van~Megen W. 1997.
\textit{Physical Review E} 55(3):3054

\bibitem{auer2001prediction}
Auer S, Frenkel D. 2001.
\textit{Nature} 409(6823):1020--1023

\bibitem{anderson2002insights}
Anderson VJ, Lekkerkerker HN. 2002.
\textit{Nature} 416(6883):811--815

\bibitem{frenkel2002playing}
Frenkel D. 2002.
\textit{Science} 296(5565):65--66

\bibitem{mirkin1996dna}
Mirkin CA, Letsinger RL, Mucic RC, Storhoff JJ. 1996.
\textit{Nature} 382(6592):607--609

\bibitem{alivisatos1996organization}
Alivisatos AP, Johnsson KP, Peng X, Wilson TE, Loweth CJ, et~al. 1996.
\textit{Nature} 382(6592):609--611

\bibitem{watson1953molecular}
Watson JD, Crick FH. 1953.
\textit{Nature} 171(4356):737--738

\bibitem{SantaLucia2004}
SantaLucia~Jr J, Hicks D. 2004.
\textit{Annual Review Of Biophysics} 33:415--440

\bibitem{park2008dna}
Park SY, Lytton-Jean AK, Lee B, Weigand S, Schatz GC, Mirkin CA. 2008.
\textit{Nature} 451(7178):553--556

\bibitem{nykypanchuk2008dna}
Nykypanchuk D, Maye MM, Van Der~Lelie D, Gang O. 2008.
\textit{Nature} 451(7178):549--552

\bibitem{macfarlane2011nanoparticle}
Macfarlane RJ, Lee B, Jones MR, Harris N, Schatz GC, Mirkin CA. 2011.
\textit{Science} 334(6053):204--208

\bibitem{casey2012driving}
Casey MT, Scarlett RT, Benjamin~Rogers W, Jenkins I, Sinno T, Crocker JC. 2012.
\textit{Nature Communications} 3(1):1209

\bibitem{rogers2015programming}
Rogers WB, Manoharan VN. 2015.
\textit{Science} 347(6222):639--642

\bibitem{wang2015crystallization}
Wang Y, Wang Y, Zheng X, Ducrot {\'E}, Yodh JS, et~al. 2015{\natexlab{a}}.
\textit{Nature Communications} 6(1):7253

\bibitem{auyeung2014dna}
Auyeung E, Li TI, Senesi AJ, Schmucker AL, Pals BC, et~al. 2014.
\textit{Nature} 505(7481):73--77

\bibitem{liu2016diamond}
Liu W, Tagawa M, Xin HL, Wang T, Emamy H, et~al. 2016{\natexlab{a}}.
\textit{Science} 351(6273):582--586

\bibitem{lewis2020single}
Lewis DJ, Zornberg LZ, Carter DJ, Macfarlane RJ. 2020.
\textit{Nature Materials} 19(7):719--724

\bibitem{zhou2024space}
Zhou W, Li Y, Je K, Vo T, Lin H, et~al. 2024.
\textit{Science} 383(6680):312--319

\bibitem{zhou2023colloidal}
Zhou W, Lim Y, Lin H, Lee S, Li Y, et~al. 2023.
\textit{Nature Materials} 23:424–--428

\bibitem{lin2017clathrate}
Lin H, Lee S, Sun L, Spellings M, Engel M, et~al. 2017.
\textit{Science} 355(6328):931--935

\bibitem{wang2022emergence}
Wang S, Lee S, Du JS, Partridge BE, Cheng HF, et~al. 2022.
\textit{Nature Materials} 21(5):580--587

\bibitem{fang2020two}
Fang H, Hagan MF, Rogers WB. 2020.
\textit{Proceedings of the National Academy of Sciences} 117(45):27927--27933

\bibitem{Hensley2022self}
Hensley A, Jacobs WM, Rogers WB. 2022.
\textit{Proceedings of the National Academy of Sciences} 119(1):e2114050118

\bibitem{he2020colloidal}
He M, Gales JP, Ducrot {\'E}, Gong Z, Yi GR, et~al. 2020.
\textit{Nature} 585(7826):524--529

\bibitem{wang2017colloidal}
Wang Y, Jenkins IC, McGinley JT, Sinno T, Crocker JC. 2017.
\textit{Nature Communications} 8(1):14173

\bibitem{ducrot2017colloidal}
Ducrot {\'E}, He M, Yi GR, Pine DJ. 2017.
\textit{Nature Materials} 16(6):652--657

\bibitem{jones2015programmable}
Jones MR, Seeman NC, Mirkin CA. 2015.
\textit{Science} 347(6224):1260901

\bibitem{rogers2016using}
Rogers WB, Shih WM, Manoharan VN. 2016.
\textit{Nature Reviews Materials} 1(3):1--14

\bibitem{seeman2017dna}
Seeman NC, Sleiman HF. 2017.
\textit{Nature Reviews Materials} 3(1):1--23

\bibitem{laramy2019crystal}
Laramy CR, O'Brien MN, Mirkin CA. 2019.
\textit{Nature Reviews Materials} 4(3):201--224

\bibitem{kahn2022designer}
Kahn JS, Gang O. 2022.
\textit{Angewandte Chemie} 134(3):e202105678

\bibitem{angioletti2016theory}
Angioletti-Uberti S, Mognetti BM, Frenkel D. 2016.
\textit{Physical Chemistry Chemical Physics} 18(9):6373--6393

\bibitem{varilly2012general}
Varilly P, Angioletti-Uberti S, Mognetti BM, Frenkel D. 2012.
\textit{The Journal of Chemical Physics} 137(9):094108

\bibitem{angioletti2013communication}
Angioletti-Uberti S, Varilly P, Mognetti BM, Tkachenko AV, Frenkel D. 2013.
\textit{The Journal of Chemical Physics} 138(2):021102

\bibitem{rogers2011direct}
Rogers WB, Crocker JC. 2011.
\textit{Proceedings of the National Academy of Sciences} 108(38):15687--15692

\bibitem{rogers2020mean}
Rogers WB. 2020.
\textit{The Journal of Chemical Physics} 153(12):124901

\bibitem{li2013thermally}
Li TI, Sknepnek R, Olvera de~la Cruz M. 2013.
\textit{Journal of the American Chemical Society} 135(23):8535--8541

\bibitem{ding2014insights}
Ding Y, Mittal J. 2014.
\textit{The Journal of Chemical Physics} 141(18):184901

\bibitem{mao2023regulating}
Mao R, Minevich B, McKeen D, Chen Q, Lu F, et~al. 2023.
\textit{Proceedings of the National Academy of Sciences} 120(52):e2302037120

\bibitem{biancaniello2005colloidal}
Biancaniello PL, Kim AJ, Crocker JC. 2005.
\textit{Physical Review Letters} 94(5):058302

\bibitem{merminod2021avidity}
Merminod S, Edison JR, Fang H, Hagan MF, Rogers WB. 2021.
\textit{Nanoscale} 13(29):12602--12612

\bibitem{Pine2022comprehensive}
Cui F, Marbach S, Zheng JA, Holmes-Cerfon M, Pine DJ. 2022.
\textit{Nature Communications} 13(1):2304

\bibitem{rogers2012reply}
Rogers WB, Crocker JC. 2012.
\textit{Proceedings of the National Academy of Sciences} 109(7):E380--E380

\bibitem{mognetti2012predicting}
Mognetti BM, Varilly P, Angioletti-Uberti S, Martinez-Veracoechea FJ, Dobnikar
  J, et~al. 2012.
\textit{Proceedings of the National Academy of Sciences} 109(7):E378--E379

\bibitem{lowensohn2019linker}
Lowensohn J, Oyarz{\'u}n B, Paliza GN, Mognetti BM, Rogers WB. 2019.
\textit{Physical Review X} 9(4):041054

\bibitem{xiong2009phase}
Xiong H, van~der Lelie D, Gang O. 2009.
\textit{Physical Review Letters} 102(1):015504

\bibitem{lowensohn2020self}
Lowensohn J, Hensley A, Perlow-Zelman M, Rogers WB. 2020.
\textit{Langmuir} 36(25):7100--7108

\bibitem{jones2010dna}
Jones MR, Macfarlane RJ, Lee B, Zhang J, Young KL, et~al. 2010.
\textit{Nature Materials} 9(11):913--917

\bibitem{tian2016lattice}
Tian Y, Zhang Y, Wang T, Xin HL, Li H, Gang O. 2016.
\textit{Nature Materials} 15(6):654--661

\bibitem{liu2016self}
Liu W, Halverson J, Tian Y, Tkachenko AV, Gang O. 2016{\natexlab{b}}.
\textit{Nature Chemistry} 8(9):867--873

\bibitem{tian2020ordered}
Tian Y, Lhermitte JR, Bai L, Vo T, Xin HL, et~al. 2020.
\textit{Nature Materials} 19(7):789--796

\bibitem{lee2022shape}
Lee S, Calcaterra HA, Lee S, Hadibrata W, Lee B, et~al. 2022.
\textit{Nature} 610(7933):674--679

\bibitem{li2023ultrastrong}
Li Y, Jin H, Zhou W, Wang Z, Lin Z, et~al. 2023.
\textit{Science Advances} 9(39):eadj8103

\bibitem{zornberg2023self}
Zornberg LZ, Lewis DJ, Mertiri A, Hueckel T, Carter DJ, Macfarlane RJ. 2023.
\textit{ACS Nano} 17(4):3394--3400

\bibitem{shani2020dna}
Shani L, Michelson AN, Minevich B, Fleger Y, Stern M, et~al. 2020.
\textit{Nature Communications} 11(1):5697

\bibitem{Hensley2023macroscopic}
Hensley A, Videb{\ae}k TE, Seyforth H, Jacobs WM, Rogers WB. 2023.
\textit{Nature Communications} 14(1):4237

\bibitem{whitelam2015statistical}
Whitelam S, Jack RL. 2015.
\textit{Annual Review of Physical Chemistry} 66:143--163

\bibitem{jacobs2016self}
Jacobs WM, Frenkel D. 2016.
\textit{Journal of the American Chemical Society} 138(8):2457--2467

\bibitem{gartner2022time}
Gartner FM, Graf IR, Frey E. 2022.
\textit{Proceedings of the National Academy of Sciences} 119(4):e2116373119

\bibitem{rogers2013kinetics}
Rogers WB, Sinno T, Crocker JC. 2013.
\textit{Soft Matter} 9(28):6412--6417

\bibitem{hurst2006maximizing}
Hurst SJ, Lytton-Jean AK, Mirkin CA. 2006.
\textit{Analytical Chemistry} 78(24):8313--8318

\bibitem{zhang2013general}
Zhang C, Macfarlane RJ, Young KL, Choi CHJ, Hao L, et~al. 2013.
\textit{Nature Materials} 12(8):741--746

\bibitem{wang2015synthetic}
Wang Y, Wang Y, Zheng X, Ducrot {\'E}, Lee MG, et~al. 2015{\natexlab{b}}.
\textit{Journal of the American Chemical Society} 137(33):10760--10766

\bibitem{oh2019colloidal}
Oh JS, Lee S, Glotzer SC, Yi GR, Pine DJ. 2019.
\textit{Nature Communications} 10(1):1--10

\bibitem{mao2020self}
Mao R, Mittal J. 2020.
\textit{The Journal of Physical Chemistry B} 124(51):11593--11599

\bibitem{zheng2023hopping}
Zheng JA, Holmes-Cerfon M, Pine DJ, Marbach S. 2023.
\textit{arXiv preprint arXiv:2310.18785}

\bibitem{lee2018modeling}
Lee-Thorp JP, Holmes-Cerfon M. 2018.
\textit{Soft Matter} 14(40):8147--8159

\bibitem{jana2019translational}
Jana PK, Mognetti BM. 2019.
\textit{Physical Review E} 100(6):060601

\bibitem{marbach2022nanocaterpillar}
Marbach S, Zheng JA, Holmes-Cerfon M. 2022.
\textit{Soft Matter} 18(16):3130--3146

\bibitem{gasser2001real}
Gasser U, Weeks ER, Schofield A, Pusey P, Weitz D. 2001.
\textit{Science} 292(5515):258--262

\bibitem{karthika2016review}
Karthika S, Radhakrishnan T, Kalaichelvi P. 2016.
\textit{Crystal Growth \& Design} 16(11):6663--6681

\bibitem{oxtoby1992homogeneous}
Oxtoby DW. 1992.
\textit{Journal of Physics: Condensed Matter} 4(38):7627

\bibitem{tkachenko2016generic}
Tkachenko AV. 2016.
\textit{Proceedings of the National Academy of Sciences} 113(37):10269--10274

\bibitem{pretti2019size}
Pretti E, Zerze H, Song M, Ding Y, Mao R, Mittal J. 2019.
\textit{Science Advances} 5(9):eaaw5912

\bibitem{landy2023programming}
Landy KM, Gibson KJ, Chan RR, Pietryga J, Weigand S, Mirkin CA. 2023.
\textit{ACS Nano} 17(7):6480--6487

\bibitem{jacobs2015self}
Jacobs WM, Frenkel D. 2015.
\textit{Soft Matter} 11(46):8930--8938

\bibitem{kravets2018plasmonic}
Kravets VG, Kabashin AV, Barnes WL, Grigorenko AN. 2018.
\textit{Chemical Reviews} 118(12):5912--5951

\bibitem{rothemund2006folding}
Rothemund PW. 2006.
\textit{Nature} 440(7082):297--302

\bibitem{wagenbauer2017we}
Wagenbauer KF, Engelhardt FA, Stahl E, Hechtl VK, St{\"o}mmer P, et~al. 2017.
\textit{ChemBioChem} 18(19):1873--1885

\bibitem{ke2014dna}
Ke Y, Ong LL, Sun W, Song J, Dong M, et~al. 2014.
\textit{Nature Chemistry} 6(11):994--1002

\bibitem{hayakawa2024}
Hayakawa D, Videb{\ae}k TE, Grason GM, Rogers WB. 2024.
\textit{ACS Nano} 18(29):19169--19178

\bibitem{videbaek2023economical}
Videb{\ae}k TE, Hayakawa D, Grason GM, Hagan MF, Fraden S, Rogers WB. 2024.
\textit{Science Advances} 10(27):eado5979

\bibitem{romano2020designing}
Romano F, Russo J, Kroc L, \ifmmode~\check{S}\else \v{S}\fi{}ulc P. 2020.
\textit{Physical Review Letters} 125(11):118003

\bibitem{russo2022sat}
Russo J, Romano F, Kroc L, Sciortino F, Rovigatti L, {\v{S}}ulc P. 2022.
\textit{Journal of Physics: Condensed Matter} 34(35):354002

\bibitem{pinto2023design}
Pinto DE, {\v{S}}ulc P, Sciortino F, Russo J. 2023.
\textit{Proceedings of the National Academy of Sciences} 120(16):e2219458120

\bibitem{liu2023inverse}
Liu H, Matthies M, Russo J, Rovigatti L, Narayanan RP, et~al. 2023.
\textit{arXiv preprint arXiv:2310.10995}

\bibitem{kahn2022encoding}
Kahn J, Minevich B, Michelson A, Emamy H, Kisslinger K, et~al. 2022.
\textit{ChemRxiv preprint https://doi.org/10.26434/chemrxiv-2022-xwbst}

\bibitem{bohlin2023designing}
Bohlin J, Turberfield AJ, Louis AA, Sulc P. 2023.
\textit{ACS Nano} 17(6):5387--5398

\bibitem{jacobs2015rational}
Jacobs WM, Reinhardt A, Frenkel D. 2015.
\textit{Proceedings of the National Academy of Sciences} 112(20):6313--6318

\bibitem{ke2012three}
Ke Y, Ong LL, Shih WM, Yin P. 2012.
\textit{Science} 338(6111):1177--1183

\bibitem{ong2017programmable}
Ong LL, Hanikel N, Yaghi OK, Grun C, Strauss MT, et~al. 2017.
\textit{Nature} 552(7683):72--77

\bibitem{Hagan2021equilibrium}
Hagan MF, Grason GM. 2021.
\textit{Reviews of Modern Physics} 93(2):025008

\bibitem{duque2023limits}
Duque CM, Hall DM, Tyukodi B, Hagan MF, Santangelo CD, Grason GM. 2023.
\textit{arXiv preprint arXiv:2309.04632}

\bibitem{Wagenbauer2017gigadalton}
Wagenbauer KF, Sigl C, Dietz H. 2017.
\textit{Nature} 552(7683):78--83

\bibitem{Sigl2021programmable}
Sigl C, Willner EM, Engelen W, Kretzmann JA, Sachenbacher K, et~al. 2021.
\textit{Nature Materials} 20(9):1281--1289

\bibitem{wei2024hierarchical}
Wei WS, Trubiano A, Sigl C, Paquay S, Dietz H, et~al. 2024.
\textit{Proceedings of the National Academy of Sciences} 121(7):e2312775121

\bibitem{Hayakawa2022}
Hayakawa D, Videb{\ae}k TE, Hall DM, Fang H, Sigl C, et~al. 2022.
\textit{Proceedings of the National Academy of Sciences} 119(43):e2207902119

\bibitem{karfusehr2024self}
Karfusehr C, Eder M, Simmel FC. 2024.
\textit{bioRxiv} :2024--02

\bibitem{Fang2022}
Fang H, Tyukodi B, Rogers WB, Hagan MF. 2022.
\textit{Soft Matter} 18(35):6716--6728

\bibitem{Videbaek2022}
Videb{\ae}k TE, Fang H, Hayakawa D, Tyukodi B, Hagan MF, Rogers WB. 2022.
\textit{Journal of Physics: Condensed Matter} 34(13):134003

\bibitem{torquato2009inverse}
Torquato S. 2009.
\textit{Soft Matter} 5(6):1157--1173

\bibitem{miskin2016turning}
Miskin MZ, Khaira G, de~Pablo JJ, Jaeger HM. 2016.
\textit{Proceedings of the National Academy of Sciences} 113(1):34--39

\bibitem{sherman2020inverse}
Sherman ZM, Howard MP, Lindquist BA, Jadrich RB, Truskett TM. 2020.
\textit{Journal of Chemical Physics} 152(14):140902

\bibitem{rechtsman2005optimized}
Rechtsman MC, Stillinger FH, Torquato S. 2005.
\textit{Physical Review Letters} 95(22):228301

\bibitem{Zeravcic2014size}
Zeravcic Z, Manoharan VN, Brenner MP. 2014.
\textit{Proceedings of the National Academy of Sciences} 111(45):15918--15923

\bibitem{halverson2013dna}
Halverson JD, Tkachenko AV. 2013.
\textit{Physical Review E} 87(6):062310

\bibitem{tkachenko2002morphological}
Tkachenko AV. 2002.
\textit{Physical Review Letters} 89(14):148303

\bibitem{lukatsky2004phase}
Lukatsky D, Frenkel D. 2004.
\textit{Physical Review Letters} 92(6):068302

\bibitem{martinez2011design}
Martinez-Veracoechea FJ, Mladek BM, Tkachenko AV, Frenkel D. 2011.
\textit{Physical Review Letters} 107(4):045902

\bibitem{scarlett2011mechanistic}
Scarlett RT, Ung MT, Crocker JC, Sinno T. 2011.
\textit{Soft Matter} 7(5):1912--1925

\bibitem{villar2009self}
Villar G, Wilber AW, Williamson AJ, Thiara P, Doye JP, et~al. 2009.
\textit{Physical Review Letters} 102(11):118106

\bibitem{hagan2006dynamic}
Hagan MF, Chandler D. 2006.
\textit{Biophysical Journal} 91(1):42--54

\bibitem{zenk2018optimizing}
Zenk J, Billups M, Schulman R. 2018.
\textit{ACS Omega} 3(12):18753--18761

\bibitem{trubiano2021thermodynamic}
Trubiano A, Holmes-Cerfon M. 2021.
\textit{Soft Matter} 17(28):6797--6807

\bibitem{chatterjee2024multi}
Chatterjee S, Jacobs WM. 2024.
\textit{arXiv preprint arXiv:2401.11234}

\bibitem{miller2010exploiting}
Miller WL, Cacciuto A. 2010.
\textit{The Journal of Chemical Physics} 133(23):234108

\bibitem{goodrich2021designing}
Goodrich CP, King EM, Schoenholz SS, Cubuk ED, Brenner MP. 2021.
\textit{Proceedings of the National Academy of Sciences} 118(10):e2024083118

\bibitem{trubiano2022optimization}
Trubiano A, Hagan MF. 2022.
\textit{Journal of Chemical Physics} 157(24):244901

\bibitem{curatolo2022assembly}
Curatolo AI, Kimchi O, Goodrich CP, Brenner MP. 2022.
\textit{bioRxiv} :2022--06

\bibitem{bolhuis2023optimizing}
Bolhuis PG, Brotzakis ZF, Keller BG. 2023.
\textit{Journal of Chemical Physics} 159(7):074102

\bibitem{whitelam2021neuroevolutionary}
Whitelam S, Tamblyn I. 2021.
\textit{Physical Review Letters} 127(1):018003

\bibitem{chennakesavalu2023unified}
Chennakesavalu S, Rotskoff GM. 2023.
\textit{Physical Review Letters} 130(10):107101

\bibitem{sanchez2018inverse}
Sanchez-Lengeling B, Aspuru-Guzik A. 2018.
\textit{Science} 361(6400):360--365

\bibitem{dijkstra2021from}
Dijkstra M, Luijten E. 2021.
\textit{Nature Materials} 20(6):762--773

\bibitem{woods2019diverse}
Woods D, Doty D, Myhrvold C, Hui J, Zhou F, et~al. 2019.
\textit{Nature} 567(7748):366--372

\bibitem{bupathy2022temperature}
Bupathy A, Frenkel D, Sastry S. 2022.
\textit{Proceedings of the National Academy of Sciences} 119(8):e2119315119

\bibitem{murugan2015multifarious}
Murugan A, Zeravcic Z, Brenner MP, Leibler S. 2015.
\textit{Proceedings of the National Academy of Sciences} 112(1):54--59

\bibitem{jacobs2021self}
Jacobs WM. 2021.
\textit{Physical Review Letters} 126(25):258101

\bibitem{huntley2016information}
Huntley MH, Murugan A, Brenner MP. 2016.
\textit{Proceedings of the National Academy of Sciences} 113(21):5841--5846

\bibitem{chen2023emergence}
Chen F, Jacobs WM. 2023.
\textit{bioRxiv} :2023--11

\bibitem{evans2024pattern}
Evans CG, O’Brien J, Winfree E, Murugan A. 2024.
\textit{Nature} 625(7995):500--507

\bibitem{van2017dissipative}
van Rossum SA, Tena-Solsona M, van Esch JH, Eelkema R, Boekhoven J. 2017.
\textit{Chemical Society Reviews} 46(18):5519--5535

\bibitem{weber2019physics}
Weber CA, Zwicker D, J{\"u}licher F, Lee CF. 2019.
\textit{Report on Progress in Physics} 82(6):064601

\bibitem{mitchison1984dynamic}
Mitchison T, Kirschner M. 1984.
\textit{Nature} 312(5991):237--242

\bibitem{spaeth2021molecular}
Sp{\"a}th F, Donau C, Bergmann AM, Kr{\"a}nzlein M, Synatschke CV, et~al. 2021.
\textit{Journal of the American Chemical Society} 143(12):4782--4789

\bibitem{nakashima2021active}
Nakashima KK, van Haren MH, Andr{\'e} AAM, Robu I, Spruijt E. 2021.
\textit{Nature Communications} 12(1):3819

\bibitem{hopfield1974kinetic}
Hopfield JJ. 1974.
\textit{Proceedings of the National Academy of Sciences} 71(10):4135--4139

\bibitem{murugan2012speed}
Murugan A, Huse DA, Leibler S. 2012.
\textit{Proceedings of the National Academy of Sciences} 109(30):12034--12039

\bibitem{zhu2023proofreading}
Zhu QZ, Du CX, King EM, Brenner MP. 2023.
\textit{arXiv preprint arXiv:2312.08619}

\bibitem{zwicker2017growth}
Zwicker D, Seyboldt R, Weber CA, Hyman AA, J{\"u}licher F. 2017.
\textit{Nature Physics} 13(4):408--413

\bibitem{wurtz2018chemical}
Wurtz JD, Lee CF. 2018.
\textit{Physical Review Letters} 120(7):078102

\bibitem{kirschbaum2021controlling}
Kirschbaum J, Zwicker D. 2021.
\textit{Journal of The Royal Society Interface} 18(179):20210255

\bibitem{cho2023nucleation}
Cho Y, Jacobs WM. 2023.
\textit{Physical Review Letters} 130:128203

\bibitem{nguyen2016design}
Nguyen M, Vaikuntanathan S. 2016.
\textit{Proceedings of the National Academy of Sciences} 113(50):14231--14236

\bibitem{das2023nonequilibrium}
Das A, Limmer DT. 2023.
\textit{Proceedings of the National Academy of Sciences} 120(40):e2217242120

\end{thebibliography}

\end{document}